# City electric power consumption forecasting based on big data & neural network under smart grid background

Zhengxian Chen[a,*], Maowei Wang[b,*], Xiao Wang[c], Conghu Li[a]

[a]*School of Computer Science & Technology, Beijing Institute of Technology, Beijing 100081, China*
[b]*Department of Electrical and Electronic Engineering, University of Nottingham, Nottingham NG7 2RD, United Kingdom*
[c]*School of Aeronautics and Astronautics, Tsinghua University, Beijing 100084, China*

**Abstract**

With the development of the electric power system, the smart grid has become an important part of the smart city. The rational transmission of electric energy and the guarantee of power supply of the smart grid are very important to smart cities, smart cities can provide better services through smart grids. Among them, predicting and judging city electric power consumption is closely related to electricity supply and regulation, the location of power plants, and the control of electricity transmission losses. Based on big data, this paper establishes a neural network and considers the influence of various nonlinear factors on city electric power consumption. A model is established to realize the prediction of power consumption. Based on the permutation importance test, an evaluation model of the influencing factors of city electric power consumption is constructed to obtain the core characteristic values of city electric power consumption prediction, which can provide an important reference for electric power related industry.

---

\* Corresponding author.
 E-mail address: 1120205042@bit.edu.cn (Zhengxian Chen), alymw23@nottingham.ac.uk (Maowei Wang)

## 1. Introduction

Smart cities are now an important topic in the field of scientific research and are closely related to people's lives. Smart city is to manage the city scientifically and conveniently by combining computer information, Internet of Things, cloud computing and other new electronic interaction and intelligent computing methods to maximize the utilization of resources and achieve a high degree of information sharing.Today's smart cities are inseparable from smart energy, and the core components of smart energy include smart electricity. Since 2008, after IBM and other large IT companies put forward the concept of smart city, it has become the consensus of the whole society to strengthen information communication, realize comprehensive information interconnection, and strengthen the application of intelligent technology in security, energy, transportation and other fields [1]. Intelligent technology has given new possibilities for city development. Smart grid is an important part of smart city and plays an important role in city construction. The relationship between them is shown in Fig. 1. The functions it provides to ensure power supply, distribute power reasonably, and save power transmission loss are of great significance for smart city construction and environmental protection. The power load or power consumption forecasting can well assist the smart grid and accurately realize these functions. Accurately predicting the city power consumption, can reasonably arrange the location and distribution of distribution stations and power plants, and at the same time coordinate the power supply as much as possible to reduce the occurrence of power outages and power cuts. Among them, big data statistics can effectively support research on electricity load and consumption, and can analyze the demand and improvement methods of smart grid from a macro perspective [2]. Various data collected in smart cities, such as total GDP, city population, etc., are of extraordinary significance.

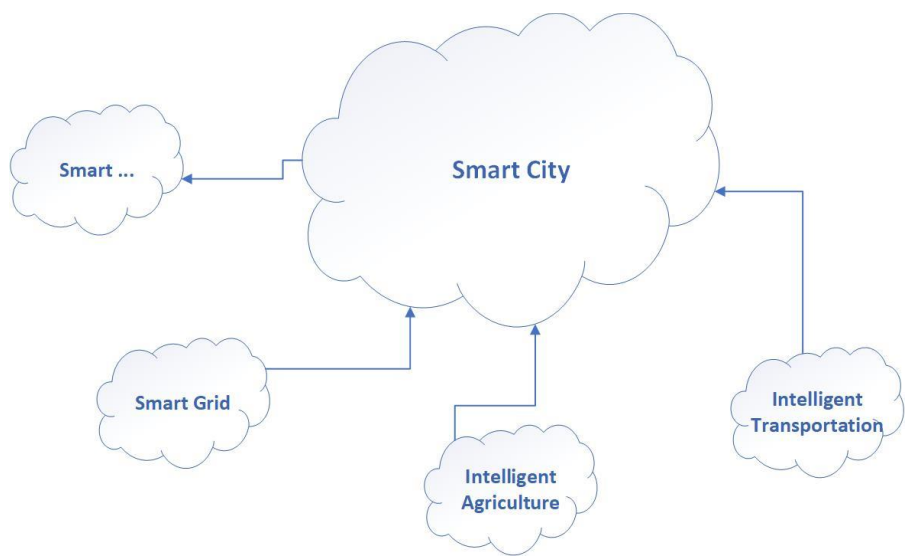

Fig. 1. Schematic diagram of Smart City.

Currently, power load and power consumption forecasting are important research topics in smart grids. Research on power forecasting has been carried out in the early 21st century, and many breakthrough achievements have been made in recent years [3-9]. Many scholars use methods derived from grey theory to forecast power load. For example, Junjie Kang et al. improved grey theory and carried out power load forecasting [10]. With the popularization of computer intelligence, methods such as deep learning and data mining are gradually adopted by people. For example, Ghulam Mohi Ud Din and others use neural networks to forecast short-term power consumption [11].

However, although most of the current researches use the latest methods, such as the use of advanced algorithms, simulated annealing, clustering algorithms, etc., when conducting power load forecasting. they only focus on how to improve the prediction accuracy by optimizing parameter data and improving model quality. The selection of data on external influencing factors is often not fully considered. Some only use several subjectively important external influencing factors for forecasting, some use a large amount of historical data, and some ignore the influence of social conditions and economic conditions. For example, in the event of a war or a large-scale epidemic, the regional power consumption obviously cannot be brought in as regular data. With the development of the economy, the factors of power consumption have also changed significantly, such as the use of electric vehicles influence electric use a lot. Therefore, combining social and economic factors, analysing from a large number of influencing factors of power consumption, and accurately determining the core influencing factors are of great significance to the prediction of power consumption. At the same time, making full use of the big data platform and selecting appropriate data for prediction is also an important guarantee for accurate prediction.

In view of this, on the basis of a large number of researches on smart grid related literature, starting from the big data collected by smart cities, by establishing a neural network model, this paper predicts the city electric power consumption in smart grids, and analyzes various influencing factors of city power consumption and the importance of each variable to find out the decisive factors affecting the city's electricity consumption. At the same time, this paper based on permutation importance test, the evaluation model of city power consumption influencing factors is proposed, which provides an important reference for related research.

**Nomenclature**

Electric Power load          Total power at a given time
Electric Power consumption       Total energy consumption over a period of time

## 2. Related concepts and research

*2.1. Definition of power load and power consumption.*

It is worth mentioning that after a detailed investigation of the papers of scholars from different countries, it is found that the definition of the term "power load" needs to be clear. Some scholars divide the "power load" into two kinds, one is the load on the power grid and the other is the load on the total energy consumption of the grid. At the same time, other scholars separate "power load" and "power consumption", and the two are parallel relations. This paper mainly studies the total energy consumption of the grid, namely the power consumption, which is made clear here. The fundamental principle of study power consumption forecasting is to analyze the main influencing factors of power load based on the existing data, and make predictions based on this. At present, scientific research institutions all over the world have carried out a lot of research on this.

Electric power consumption forecasting is the focus of this paper. Power consumption can be divided into short-term power consumption forecast and long-term power consumption forecast according to the forecast time [12]. There is no obvious boundary between the two. The annual city electric consumption performed in this paper is part of the long-term power consumption forecast.

*2.2. Artificial neural network.*

Artificial neural network is a model of machine learning. Neural network can be used in prediction, pattern recognition, clustering, classification and other purposes, and has become a common tool in many disciplines in recent years. For example, there are applications of neural networks in artificial intelligence, computer science, information security, statistics, and other fields. Neural networks are derived from bionics to a certain extent and can perform distributed parallel information processing. Biological neural network is the brain neural network of animals and humans. It has many neurons, which can be connected to each other and form a highly complex dynamic network. Artificial neural networks imitate biological neural networks, and perform intelligent operations by designing related models and algorithms.

The principle of the neural network can be simply represented by the neuron model in Fig. 2. Neurons are the basic components of the neural network. The neuron model includes functions such as output, input, and calculation. It imitates the dendrites, nucleus, axon etc. [13-15]

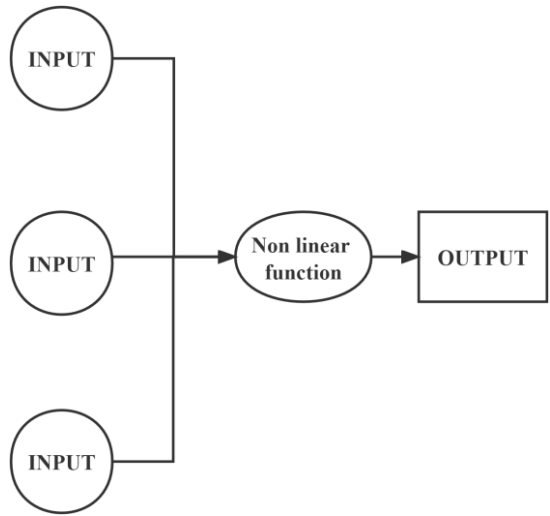

Fig. 2. Neurons.

## 3. Research theory of city electric power consumption prediction

*3.1. Influencing factors and forecasting analysis of city electric power consumption*

As mentioned above, city electric power consumption is a hot research topic in smart grid area, which is closely related to power grid planning. As for the influencing factors and forecasting analysis of city electric power consumption, related research has gradually emerged. Liang Chaohui et al. put forward in the paper that city industrial scale, urbanization degree, built-up area scale, city population scale and other factors are the main influencing factors of city electric power consumption [16]. Carlos Pena-Guzman et al. mentioned above used the number of residential users, total energy bill, natural gas bill, user income, and GDP as variables to make predictions.

It can be seen that the variables used in the relevant research are all city macro data, but the number is limited, mostly ranging from 5-12. Due to the complexity of city electric power consumption forecasting, fewer variables may not fully cover the influencing factors of city electric power consumption. On this basis, this paper proposes an analysis model for the influencing factors of city electric power consumption, and selects 85 items in the city macro data as the influencing factors (eigenvalues) of city electric power consumption for analysis. And innovatively divide it, put forward the classification standard of city electric power consumption influencing factors, and divide 85 city macro data into two categories: core influencing factors and common (potential) influencing factors of city electric power consumption. The core influence of city electric power consumption will be studied in detail later. The specific definition, analysis and division are as follows:

Table 1. City data division and definition.

| Date type | Core influencing factors | Common influencing factors |
|---|---|---|
| Definition | The core influencing factors need to have a strong relationship with urban electricity consumption, which can objectively and accurately reflect the main characteristics of the city. | Common influencing factors may not have a strong relationship with city electric power consumption, but they can logically affect city electric power consumption to a certain extent. |

According to the above table, 10 of the 85 city data are selected as core variables, as shown in the following table:

Table 2. City data Core influencing factors

| Data name | Unit | Illustration |
|---|---|---|
| Gross regional domestic product (GDP) | 10000CNY | The most important parameter reflects the size and development of a city |
| Total population | 10000 | City population |
| Highway passenger volume | 10000 | Road passenger volume reflects whether a city is a transportation hub or not, as well as the greater demand for electricity due to the popularity of electric vehicles |
| Highway freight volume | 10000 ton | Road freight volume, reflecting whether the city is a transportation hub, and due to the popularity of electric vehicles, the demand for electricity is greater |
| Number of domestic enterprises | 1 | Number of local enterprises, reflecting commercial and industrial electricity consumption |
| Number of employees on the job | 10000 | The number of employees reflects the size of a city and its economic development |
| Total telecom business | 10000CNY | Total income of telecommunication service, have certain direct relation with electric power demand |
| number of mobile phone users | 10000 | The number of mobile phone users, which reflects the overall use of electrical appliances in residents and reflects the residential electricity consumption |
| area of land | Square kilometers | City size |
| Number of industrial enterprises above designated size | 1 | In response to the number of large industrial enterprises in the city, it is closely related to industrial electricity consumption |

The remaining 75 variables were taken as common influencing factors and included in the neural network analysis, but the importance test and analysis below will not include them. See Table 3 for details, which omit individual variables with strong linear relationship.

Table 3. City data common influencing factors

| Data name | Unit | Data name | Unit |
| --- | --- | --- | --- |
| Births | 1 | Total profit | 10000CNY |
| Deaths | 1 | Number of doctors | 1 |
| Natural Growth Rate of Population | Permillage | Number of beds in hospitals and health centers | 1 |
| Number of employees in transportation, storage, post and telecommunications industry | 10000 | Number of hospitals and health centers | 1 |
| Number of employees in accommodation and catering industry | 10000 | Number of Museums | 1 |
| Number of employees in information transmission, computer service and software industry | 10000 | Number of invention patents authorized | 1 |
| Number of employees in public management and social organizations | 10000 | Total wages of on-the-job employees | 10000CNY |
| Number of employees in agriculture, forestry, animal husbandry and fishery | 10000 | Average number of on-the-job employees | 10000 |
| Number of employees in manufacturing industry | 10000 | General budgetary expenditure of local finance | 10000CNY |
| Number of employees in health, social insurance and social welfare | 10000 | General budgetary revenue of local finance | 10000CNY |
| Number of employees in resident service and other service industries | 10000 | Number of urban employees participating in basic old-age insurance | 1 |
| Value added of the secondary industry | 10000CNY | Number of Foreign-invested Enterprises | 1 |
| Number of employees in construction industry | 10000 | Number of people participating in unemployment insurance | 1 |
| Number of employees in the real estate industry | 10000 | Number of full-time teachers in primary schools | 1 |
| Number of employees in wholesale and retail trade | 10000 | Number of primary schools | 1 |
| Number of employees in the education industry | 10000 | Number of primary school students | 10000 |
| Number of employees in culture, sports and entertainment | 10000 | Total number of households at the end of the year | 10000 |
| Number of employees in | 10000 | Balance of deposits of | 10000CNY |
| water conservancy, environment and public facilities management | | financial institutions at the end of the year | |
| Number of employees in electricity, gas and water production and supply industry | 10000 | Education expenditure | 10000CNY |
| Number of employees in scientific research, technical services and geological exploration | 10000 | Number of full-time teachers in ordinary middle schools | 1 |
| Number of employees in leasing and commercial service industry | 10000 | Number of students in ordinary middle schools | 10000 |
| Number of employees in the financial industry | 10000 | Number of ordinary secondary schools | 1 |
| Number of employees in the primary industry | 10000 | Number of ordinary colleges and Universities | 1 |
| Proportion of employees in the primary industry | percentage | Total water resources | 10000 cubic meters |
| Number of employees in the secondary industry | 10000 | Total retail sales of consumer goods | 10000CNY |
| Proportion of employees in the secondary industry | percentage | Scientific expenditure | 10000CNY |
| Number of employees in the tertiary industry | 10000 | Average salary of employees | 1CNY |
| Proportion of employees in the tertiary industry | percentage | Export volume of goods | 10000CNY |
| Number of patent authorizations | 1 | Import volume of goods | 10000CNY |
| Number of patent applications | 1 | Total postal services | 10000CNY |
| Number of students in secondary vocational and technical schools | 1 | Total sales of wholesale and retail trade above Designated Size | 10000CNY |
| Number of full-time teachers in secondary vocational education schools | 1 | Year end balance of urban and rural residents' savings | 10000CNY |
| Number of secondary vocational education schools | 1 | Balance of various loans of financial institutions at the end of the year | 10000CNY |
| Total collection of books in Public Libraries | 1000 | Per capita gross regional product | 1CNY |
| Value added of the primary industry | 10000CNY | Growth rate of gross regional product | percentage |
| Value added of the primary industry | 10000CNY | Value added of the tertiary industry | 10000CNY |

A total of 85 kinds of city data were input into the neural network model as eigenvalues of city electricity power consumption.

*3.2. Selection principles of city data*

In order to make the data as accurate as possible, this paper selects 269 cities with complete data of 85 items for research, and does not use any estimation and filling methods. At the same time, considering the uncertain impact of the COVID-19 on city production and residents' activities, this paper uses the data of 269 cities before 2020 to eliminate the uncertainty caused by the COVID-19.

*3.3. Construction of city electricity power consumption neural network model*

The relationship between city electricity power consumption and its influencing factors is essentially equivalent to a neural network, as shown in the Fig. 3. Among them, 85 city data, as independent variables, have an impact on city electricity power consumption. Each city data is equivalent to a neuron, and different city data also have a certain impact. For example, the highway passenger volume mentioned above has a certain nonlinear relationship with the city population. Telecom business volume, with the Internet penetration rate, the number of mobile phone users have a relationship. The data of 85 cities were input into the neural network system as eigenvalue variables, and the neural network model of city electricity power consumption was established through machine learning to realize the prediction and deployment control of city electricity in advance.

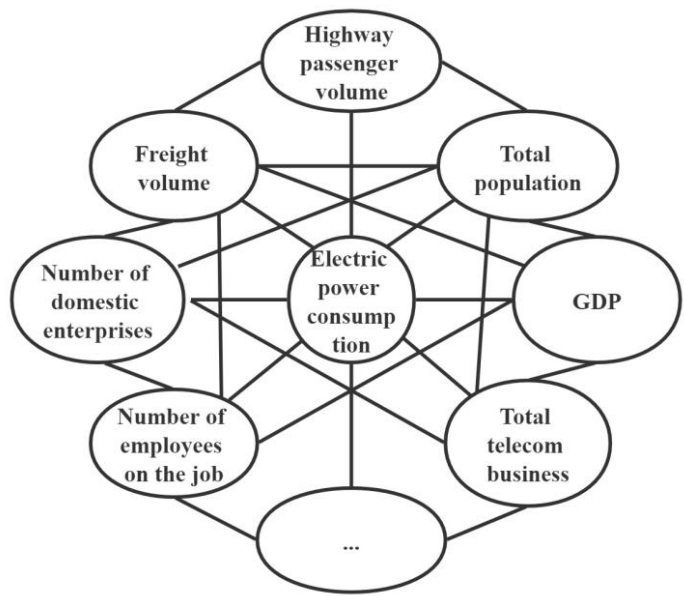

Fig. 3. Neural network.

The mapping table between input and output was sorted out. The output value was city electricity power consumption, and the input value was 85 cities data to build a neural network learning model, as shown in Fig. 4.

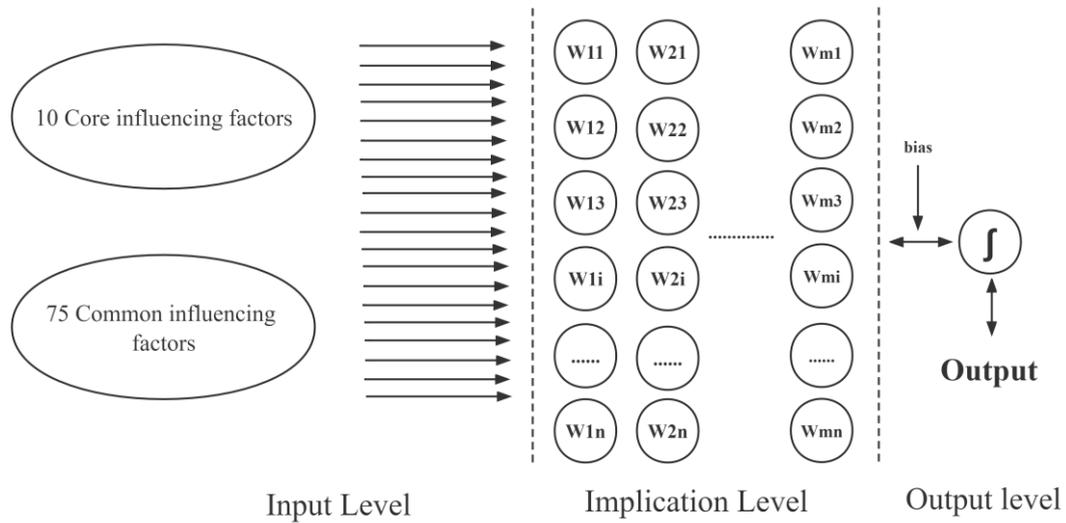

Fig. 4. Input and Output.

### 3.4. City electric power consumption prediction data division

In the neural network learning model, the data of 269 cities are divided into four groups, training set, validation set, test set A and test set B. Due to the large differences between cities, in order to accurately and comprehensively test the accuracy of the neural network model, a total of 49 cities of different sizes, types were manually selected as test set B. At the same time, the data of the remaining 220 cities are randomly assigned to the training set, validation set and test set A according to the ratio of 92:4:4. In this way, both randomness and scientificity can be taken into account in the two test sets. The final model evaluation is carried out according to the results of test set B, and test set A is used as reference. See the table below for specific division.

Table 4. City data Division.

| Data Group | Numbers of Cities Data |
|---|---|
| Training set | 202 |
| Validation set | 9 |
| Test set A | 9 |
| Test set B | 49 |

### 3.5. City electric power consumption prediction, control and analysis process based on neural network

The city electric power consumption prediction, control and analysis process based on neural network are shown in Fig. 5. The detailed process and specific application method of this model are given in this paper.

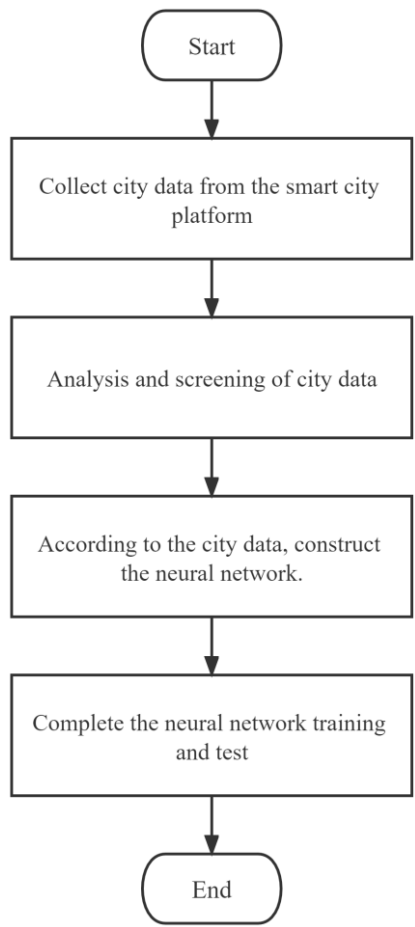

Fig. 5. Prediction, control and analysis process.

## 4. Model training and prediction results

### 4.1. Neural network training and testing

The neural network is built based on MATLAB, and relevant parameters is set: the maximum number of training times is 1000, the learning rate was 0.01, the number of hidden layer nodes is 10, and the transfer function is Purelin and tansig function. After post-test, Tansig activation function performs better.

The following figure shows its network model.

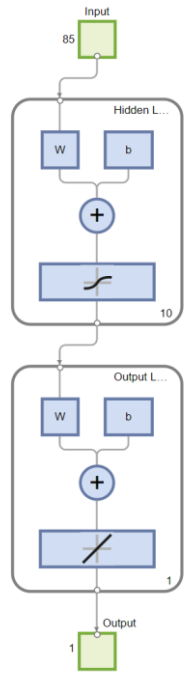

Fig. 6. Neural Network.

Tansig Function is Symmetric sigmoid transfer function:

$$tansig(x) = \frac{2}{1+e^{-2x}} - 1 \tag{1}$$

Purelin Function is Linear transfer function.

$$y = x \tag{2}$$

The training results are shown in the Fig. 7. After several training process, the coefficient of determination ($R^2$) of the test set B is around 0.81 – 0.89, select the best results for the regression, performance display.

The comparison of the predicted and actual values of test set B are generally in line with the requirements. The test results of the model under three different training are given, it shown that the stability of the model is good:

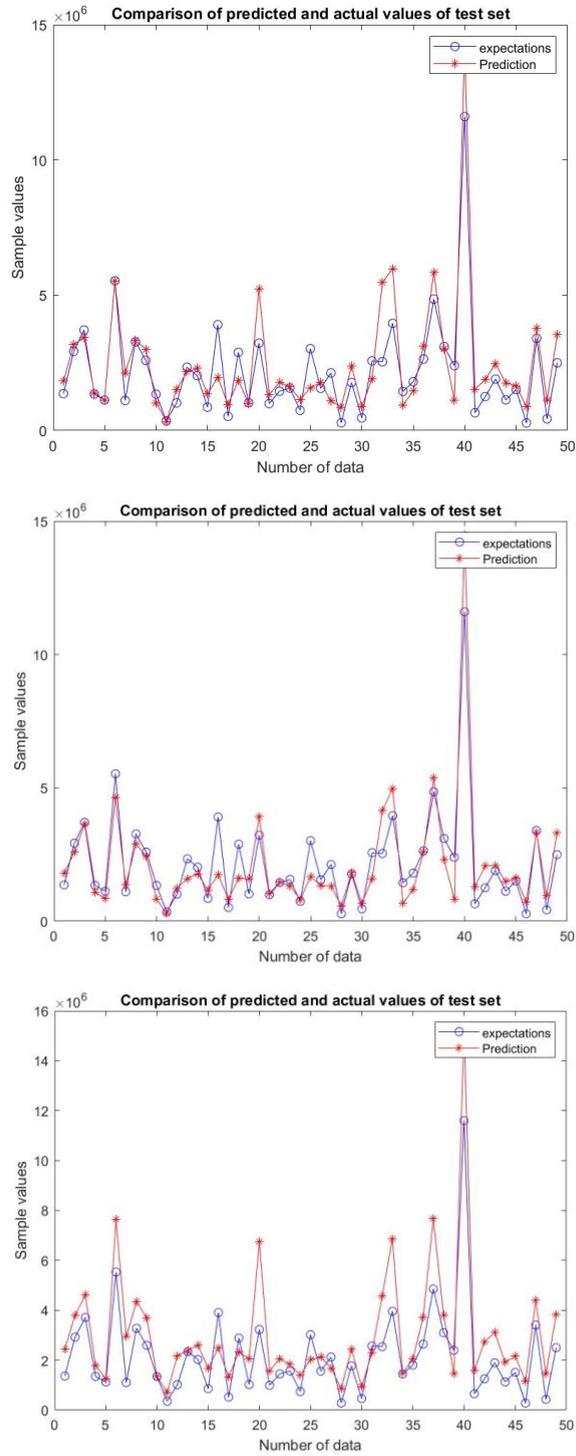

Fig. 7. Comparison of predicted and actual values of test set B.

The best regression train plots is shown in Fig 8.

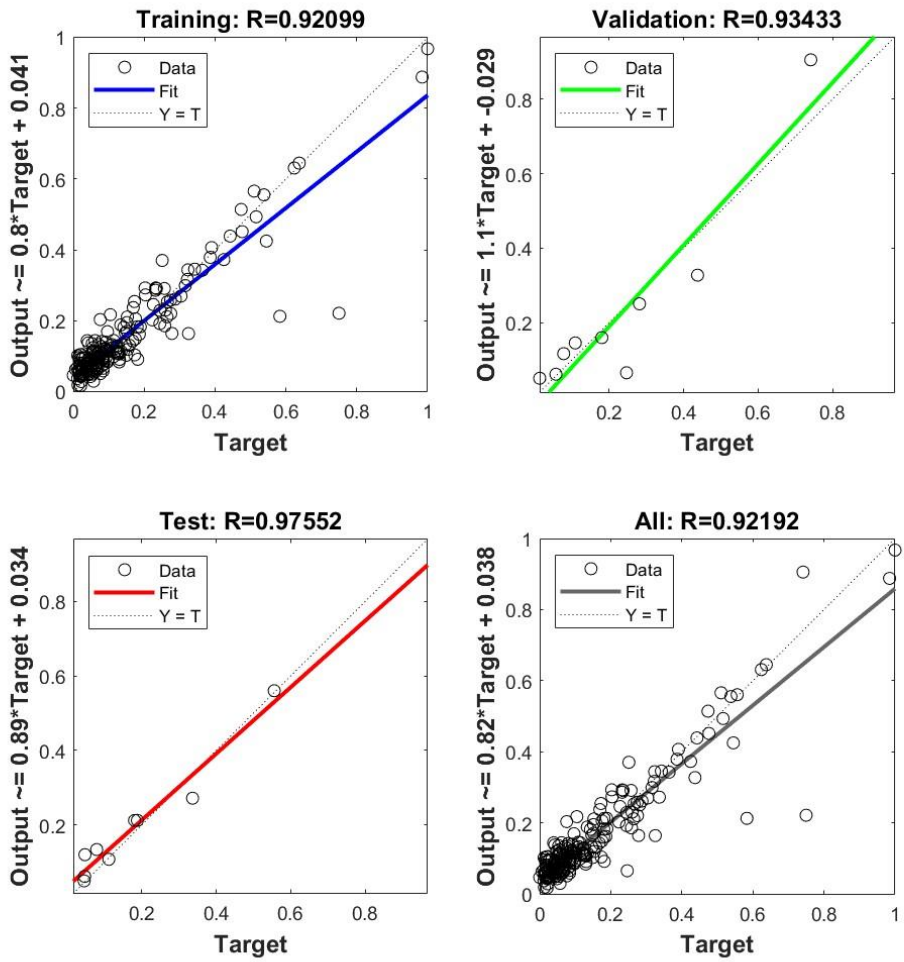

Fig. 8. Regression training plots.

The performance is shown in Fig 9.

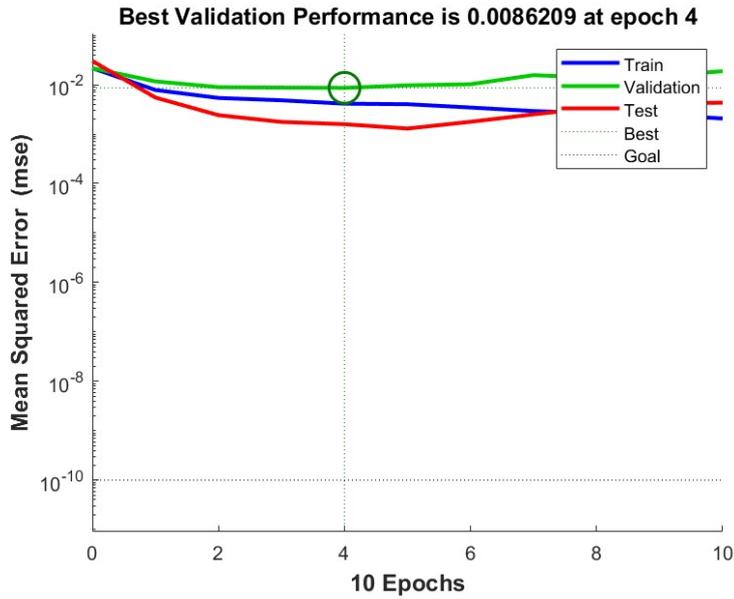

Fig. 9. Best validation performance.

The training state is shown in Fig. 10.

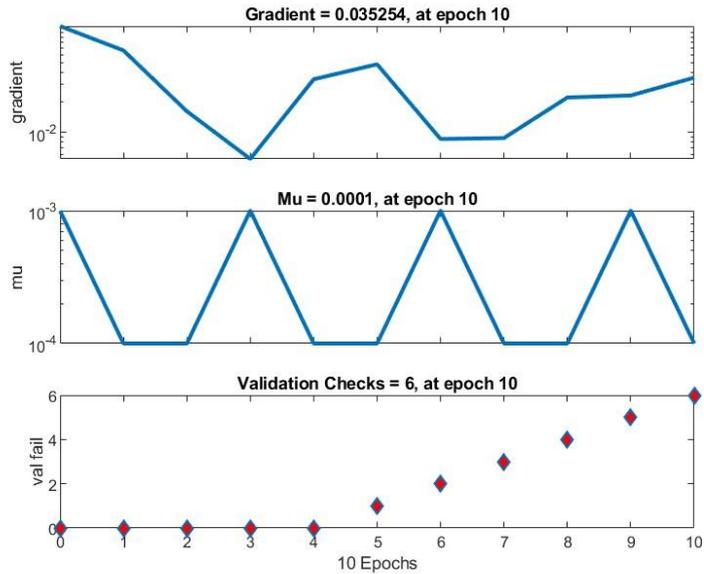

Fig. 10. Train state.

## 4.2. Results analysis

$R^2$ coefficient, also known as coefficient of determination, is defined as:

$$R^2 = \frac{SSR}{SST} = 1 - \frac{SSE}{SST} \tag{3}$$

SSR is the sum of squares of regressions, SSE is the sum of squares of residuals, and SST is the total sum of squares.
- $R^2 = 1$ : Ideally, all predicted values are equal to true values.
- $R^2 = 0$ : Simply predict that all y values are equal to the mean of y. Or other special case.
- $R^2 < 0$ : The prediction ability of the model is poor.

In the prediction above, the $R^2$ value is 0.9, the average absolute error MAE is 664K, and the root mean square error RMSE is 932K. It can be seen from the above figure that the prediction can basically predict the power consumption relatively accurately. The overall model performs well in small and medium-sized cities, and the overall predicted value is slightly larger than the actual value in large cities. It proves that the model is more suitable for small and medium-sized cities.

R coefficient, the full name of correlation coefficient, represents the correlation between predicted output and target output. The closer R value is to 1, the closer the relationship between prediction and output data is, and the closer R value is to 0, the greater the randomness of the relationship between prediction and output data is.

It is defined as:

$$R = \frac{Cov(X,Y)}{\sqrt{Var[X]Var[Y]}} \tag{4}$$

Where, $Cov(X,Y)$ is the covariance, $Var[X]$, $Var[Y]$ is the variance. Due to lack of space, the relevant definitions will not be described here.

It can be seen that the $R$ value of training set and test set is more than 0.9, and the training effect is in line with the requirements

## 5. Importance analysis of eigenvalues

### 5.1. Permutation Importance

How to determine the importance of eigenvalues is an important issue in neural network research. The PI method is suitable for tabular data, and its evaluation of feature importance depends on the degree to which the model performance score decreases after the feature is randomly rearranged. Randomly rearrange an eigenvalue, construct a tainted dataset, and then compute the performance of the trained model on that dataset. The importance of this feature value is obtained by observing how much its score decreases. The importance scores of eigenvalues are written as:

$$S = K - \frac{1}{L\sum_{l=1}^{L} K} \tag{5}$$

S is the importance; K is the Performance score; L is the times of experiment;

The method has been applied by related scholars in the selection of short-term power load forecast variables. Nantian Huang et al. used this method to study 243 eigenvalues in power load forecasting [17]. This method can also be used in the study of city electricity consumption. In this paper, PI test is carried out on 10 main variables, and their permutation important values are obtained and analyzed.

Aiming at the problem of city electricity power consumption prediction in this paper, the model is slightly adjusted, and all parameters are not studied. At the same time, the model scoring basis is set as the value of $R^2$, and an evaluation model for the influencing factors of city power consumption is proposed.

$$S = \frac{\sum_{L=1}^{L}\sqrt{(R^2-R_0{}^2)^2}}{L} \tag{6}$$

L is the times of experiment;
The result is shown in the table below:

Table 5. Permutation importance test result

| Data name | PI Score |
|---|---|
| Gross regional domestic product (GDP) | 0.15 |
| Total population | 0.19 |
| Highway passenger volume | 0.47 |
| Highway freight volume | 0.33 |
| Number of domestic enterprises | 0.09 |
| Number of employees on the job | 0.49 |
| Total telecom business | 0.58 |
| number of mobile phone users | 0.29 |
| area of land | 0.41 |
| Number of industrial enterprises above designated size | 0.38 |

It can be seen from the above table that the total amount of telecommunication business is the most important for the prediction of city power consumption, followed by the number of employees, road passenger volume, and land area, followed by GDP, population, highway freight volume, number of mobile phone users, and number of industrial enterprises above designated size. The number of local firms barely affected the neural network's operation.

The interpretation of this result is as follows: the total amount of telecommunication business can be accurately counted in smart cities and is directly related to power consumption. Although GDP and population are important characteristics of cities, and they are related to electricity consumption to a certain extent, but they are not absolutely related. For example, many cities with agriculture and mining as the main industries, most of which constitute GDP, their main energy sources are oil, coal mines, etc., and the power consumption is not too high.

Of course, the evaluation model has a certain correlation with the actual situation, but since there are as many as 85 eigenvalues input by the neural network, and many variables have a certain nonlinear relationship, the evaluation results are not very accurate and can only be used as a reference. For neural networks with fewer eigenvalues, the PI test will give better results.

## 6. Conclusion

Under the background of smart city and smart grid, this paper proposes a neural network prediction model for city electric power consumption. Through the input of 85 eigenvalues, the prediction analysis of city electric power consumption is carried out. The prediction model takes into account the influence of many nonlinear factors on city electric power consumption, and the prediction can basically reflect the real situation. At the same time, the paper clarifies the conceptual definitions of power load and power consumption, and proposes an evaluation model for the influencing factors of city power consumption based on the PI test, which provides a reference for related research. This model is more suitable for the case of fewer eigenvalue inputs, and more correlated eigenvalue inputs will bring certain interference. The models proposed in this paper need to be combined with the actual situation if it be used in real production, and they still have room for improvement. Future research will also focus on improving the accuracy and scientificity of the models.

# References

[1] Su, Kehua, Jie Li, and Hongbo Fu. (2011) "Smart city and the applications." *2011 international conference on electronics, communications and control (ICECC)*. IEEE, 2011.

[2] Fang, X., Misra, S., Xue, G., & Yang, D. (2011). "Smart grid—The new and improved power grid: A survey. " *IEEE communications surveys & tutorials*, 14.4 (2011), 944-980.

[3] Camero, Andrés, and Enrique Alba. (2019) "Smart City and information technology: A review." *cities* 93 (2019): 84-94.

[4] Bakıcı, Tuba, Esteve Almirall, and Jonathan Wareham. (2013) "A smart city initiative: the case of Barcelona." *Journal of the knowledge economy* 4.2 (2013): 135-148.

[5] Yan-e, Duan. (2011) "Design of intelligent agriculture management information system based on IoT." *2011 Fourth International Conference on Intelligent Computation Technology and Automation*. Vol. 1. IEEE, 2011.

[6] Dileep, G. (2020) "A survey on smart grid technologies and applications." *Renewable energy* 146 (2020): 2589-2625.

[7] Moslehi, Khosrow, and Ranjit Kumar. (2010) "A reliability perspective of the smart grid." *IEEE transactions on smart grid* 1.1 (2010): 57-64.

[8] Ma, Ruofei, et al. (2013) "Smart grid communication: Its challenges and opportunities." *IEEE transactions on Smart Grid* 4.1 (2013): 36-46.

[9] Tuballa, Maria Lorena, and Michael Lochinvar Abundo. (2016) "A review of the development of Smart Grid technologies." *Renewable and Sustainable Energy Reviews* 59 (2016): 710-725.

[10] Kang, Junjie, and Huijuan Zhao. (2012) "Application of improved grey model in long-term load forecasting of power engineering." *Systems Engineering Procedia* 3 (2012): 85-91.

[11] Peña-Guzmán, Carlos, and Juliana Rey. (2020) "Forecasting residential electric power consumption for Bogotá Colombia using regression models." *Energy Reports* 6 (2020): 561-566.

[12] Xia, Changhao, Jian Wang, and Karen McMenemy. (2010) "Short, medium and long term load forecasting model and virtual load forecaster based on radial basis function neural networks." *International Journal of Electrical Power & Energy Systems* 32.7 (2010): 743-750.

[13] Abiodun, Oludare Isaac, et al. (2018) "State-of-the-art in artificial neural network applications: A survey." *Heliyon* 4.11 (2018): e00938.

[14] Wang, Sun-Chong. (2003) "Artificial neural network." *Interdisciplinary computing in java programming*. Springer, Boston, MA, 2003. 81-100.

[15] Albawi, Saad, Tareq Abed Mohammed, and Saad Al-Zawi. (2017) "Understanding of a convolutional neural network." *2017 international conference on engineering and technology (ICET)*. Ieee, 2017.

[16] Liang, Chaohui. (2010) " The Impact Factors of Power Consumption of Chinese Cities ——— Based on City-level Panel Data Analysis." Shanghai Journal of Economic 2010(07):22-30.

[17] Huang, Nantian, Guobo Lu, and Dianguo Xu. (2016)"A permutation importance-based feature selection method for short-term electricity load forecasting using random forest." *Energies* 9.10 (2016): 767.